\documentclass[12pt]{iopart}
\usepackage[english]{babel}
\usepackage{amsfonts}
\usepackage{amssymb}
\usepackage{graphicx}
\usepackage{amsthm}
\usepackage{bm}

\begin{document}

\title[Coulomb excitation of $^{229m}$Th by muons]{Cross section of the Coulomb excitation of $^{229m}$Th by low energy muons.}

\author{E.~V.~Tkalya}

\address{P.~N.~Lebedev Physical Institute of the Russian
Academy of Sciences, 119991, 53 Leninskiy pr., Moscow, Russia}

\address{National Research Nuclear University MEPhI, 115409,
Kashirskoe shosse 31, Moscow, Russia}

\address{Nuclear Safety Institute of RAS, Bol'shaya Tulskaya
52, Moscow 115191, Russia}

\ead{tkalya\_e@lebedev.ru}

\vspace{10pt}

\begin{abstract}
The inelastic scattering cross section for muons, $\mu^-$,  with
energies $E$=9--100~eV from the $^{229}$Th nuclei is calculated in the framework of the second order of the perturbation theory for the
quantum electrodynamics. The dominant contribution to the excitation of the low energy isomer $^{229m}$Th$(3/2^+,8.19\pm0.12$~eV) comes from the $E2$ multipole. The excitation cross section reaches the value of $10^{-21}$ cm$^2$ in the range $E\approx$10~eV. This is four to five orders of magnitude larger than the electron excitation cross section and enough for efficient excitation of $^{229m}$Th on the muon beam at the next generation of muon colliders.
\end{abstract}


\maketitle

\section{Introduction}
\label{sec:Introduction}

Great interest in the low-lying isomeric state
$3/2^+(E_{\rm{is}}<10$~eV) in the $^{229}$Th nucleus is caused
by the possibility of designing the ultra-precise nuclear clock
\cite{Peik-03,Rellergert-10,Campbell-12,Peik-15}, the nuclear
laser in the optical range \cite{Tkalya-11,Tkalya-13}, and the
nuclear light-emitting diode of the VUV range
\cite{Tkalya-20-PRL}, as well as the study of a number of unusual
processes: excitation and decay of $^{229m}$Th by laser radiation
through the electron shell at the electron bridge
\cite{Strizhov-91,Tkalya-92-JETPL,Tkalya-92-SJNP,Kalman-94,Tkalya-96,Porsev-10-PRL,
Muller-19, Borisyuk-19-PRC}, control of the isomeric level
$\gamma$ decay via the boundary conditions \cite{Tkalya-18-PRL} or
chemical environment \cite{Tkalya-00-JETPL,Tkalya-00-PRC},
$\alpha$ decay of the $^{229m}$Th isomer \cite{Dykhne-96} and
accompanying bremsstrahlung \cite{Tkalya-99-PRC}, the relative
effects of the variation of the fine structure constant and the
strong interaction parameter
\cite{Flambaum-06,Litvinova-09,Berengut-09}, a check of the
exponentiality of the decay law at long times \cite{Dykhne-98} and
others.

The $^{229m}$Th isomeric state has the lowest excitation energy
among all known nuclei. According to the latest data
\cite{Peik-20} its energy, $E_{\rm{is}}$, is $8.19\pm 0.12$~eV.
This result is close to the value $E_{\rm{is}}=8.28 \pm 0.17$~eV
obtained in Ref.~\cite{Seiferle-19}, to the measurements of
Ref.~\cite{Sikorsky-20} $E_{\rm{is}}=8.10 \pm 0.17$~eV, and
$E_{\rm{is}}=7.8\pm 0.5$~eV from Ref.~\cite{Beck-07}. Prior to
that, for a fairly long period of time from 1990 to 2007, it was
believed that $E_{\rm{is}}<5$~eV \cite{Reich-90,Helmer-94}.

Currently, the $\alpha$ decay of $^{233}$U is practically the only
way to obtain the $^{229m}$Th isomer. Effective excitation of
$^{229m}$Th by laser radiation is not feasible today, since it
requires knowledge of the transition energy with a much greater
accuracy than that achieved now. Therefore, in the work
\cite{Tkalya-20-PRL}, it was proposed to excite $^{229m}$Th by
inelastic electron scattering. It turned out that in the beam
energy region $E\approx10$~eV, the excitation cross section
reaches the value of $10^{-25}$~cm$^2$. Such a large cross section
indicates that the method of obtaining of $^{229m}$Th using beams
of negatively charged particles is promising. As a continuation of
the work \cite{Tkalya-20-PRL}, we consider here the process of
inelastic scattering of low-energy muons from the $^{229}$Th
nuclei.

A prerequisite for such work may be the following considerations.
In the Born approximation, cross sections for nuclear excitation
to an isomeric state with the energy $E_{\rm{is}}$ were obtained
analytically in Ref.~\cite{Alder-56} for the electric excitation
and in Ref.~\cite{Tkalya-12-PRC-Born} for the magnetic excitation.
The cross sections for the magnetodipole ($M1$) transitions and
electroquadrupole ($E2$) transitions have the form
\cite{Alder-56,Tkalya-12-PRC-Born}
%
% Eq. Born M1 and E2 Cross Sections
%
%
\begin{eqnarray}
\sigma^{\rm{Born}}_{M1}&=& \frac{16\pi^2}{9} e^2
\frac{2E-E_{\rm{is}}}{E}
\ln{\frac{\sqrt{E}+\sqrt{E-E_{\rm{is}}}}{\sqrt{E}-\sqrt{E-E_{\rm{is}}}}} B(M1),\nonumber \\
\sigma^{\rm{Born}}_{E2}&=& \frac{64\pi^2}{225} e^2 m^2
\sqrt{\frac{E-E_{\rm{is}}}{E}} B(E2), \nonumber
\end{eqnarray}
where $E=p^2/2m$ is the kinetic energy of a scattering particle
with the mass $m$ and momentum $p=|{\bf{p}}|$ in the initial state
in the nonrelativistic approximation, $B(M1)$ and $B(E2)$ are the
reduced probabilities of the nuclear transition from the ground
state to the isomeric state.

As follows from the above formulas, the magnetic dipole cross
section, $\sigma^{\rm{Born}}_{M1}$, does not depend on the
parameters of the scattered particle and is the same for electrons
and muons. On the other hand, the cross section
$\sigma^{\rm{Born}}_{E2}$ includes the square of the mass of the
scattered particle \cite{Alder-56,Tkalya-12-PRC-Born}. Since the
masses of the muon, $m_{\mu}$, and electron, $m_e$, are related as
$m_{\mu}/m_e\approx 206.77$, then in the scattering of a muon and
an electron of the same energy we have
$\sigma^{\rm{Born}}_{E2}(\mu^-)/\sigma^{\rm{Born}}_{E2}(e^-)\approx
4\times 10^4$. If this relation is retained on going to the
low-energy region, then we can expect that the total $M1+E2$ cross
section for scattering of muons by the $^{229}$Th nuclei will be
several orders of magnitude larger than the cross section for
scattering of slow electrons considered in \cite{Tkalya-20-PRL}.

In this paper, we will consider the $^{229}$Th excitation by
muons, calculate the cross section of the
$^{229}$Th$(\mu^-,{\mu^-})$$^{229m}$Th reaction for the low energy
muons (show that it is indeed much larger than the cross section
of the $^{229}$Th$(e^-,{e^-})$$^{229m}$Th reaction), and discuss
the new possibilities for studying the properties of the
$^{229}$Th nuclear transition $5/2^+(0.0)\leftrightarrow
3/2^+(8.10\pm0.17$~eV) at the next generation of the muon
colliders.

\section{Cross section}
\label{sec:CS}

\begin{figure}
 \includegraphics[angle=0,width=0.35\hsize,keepaspectratio]{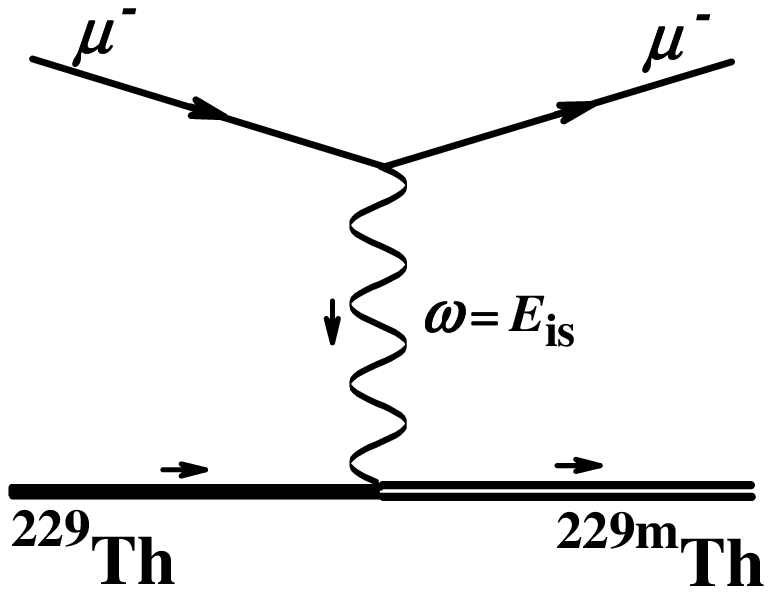}
 \caption{The Feynman diagram of the inelastic scattering process.}
 \label{fig:FeynmanDiagram}
\end{figure}

The cross section of the nuclear excitation by muons (electrons)
in the process shown in Fig.~\ref{fig:FeynmanDiagram}
%
%  Figure 1. Feynman Diagram
%
was obtained in the framework of quantum electrodynamics in the
work \cite{Tkalya-20-PRL}. In the energy region where the kinetic
energy of the muon in the initial state, $E={\cal{E}}-m_{\mu}$
(${\cal{E}}$ is the relativistic energy
${\cal{E}}=\sqrt{{\bf{p}}^2+m_{\mu}^2}$) satisfies the condition
$E \ll m_{\mu}$, the cross section is given by
%
% Eq. Cross Section (Non Relativistic)
%
\begin{eqnarray}
\fl \sigma_{E(M)L} = 4e^2\lambda^2_{\gamma_{\rm{is}}}
\left(\frac{E}{E_{\rm{is}}}\right)^{-3/2}
\left(\frac{E}{E_{\rm{is}}}-1\right)^{-1/2} \frac{B(E(M)L;J_i\rightarrow{}J_f)}{a_B^{2L}}\times \nonumber\\
    \sum_{l_i,j_i \atop{}l_f,j_f}
\frac{(2l_i+1)(2j_i+1)(2j_f+1)}{(2L+1)^2} \left(C^{l_f0}_{l_i0L0}\right)^2 \left\{
\begin{array}{ccc}
 l_i & L & l_f \\
 j_f &1/2& j_i
\end{array}
\right\}^2
\left|\tilde{{\rm{\textsl{m}}}}^{E(M)L}_{fi}\right|^2,
\label{eq:CS_NR}
\end{eqnarray}
where $\lambda_{\gamma_{\rm{is}}}=2\pi/E_{\rm{is}}$ is the
wavelength of the isomeric nuclear $\gamma$ transition, $a_B$ is
the Bohr radius, $a_B=1/(e^2m_e)$, $e$ and $m_e$ are the charge
and mass of the electron, $j_{i,f}$ and $l_{i,f}$ are the total
and orbital angular momenta of the muon in the initial and final
states. In the case of the $ML$ transition, one replaces $l_i$ by
$l'_i=2j_i-l_i$. The muon matrix elements in Eq.~(\ref{eq:CS_NR})
are
%
% Muon Matrix Elements (small Eis)
%
\begin{equation}
\eqalign{
\tilde{{\rm{\textsl{m}}}}_{fi}^{EL} = \int_0^{\infty}
[g_i(x)g_f(x)+ f_i(x)f_f(x)]\frac{dx}{x^{L-1}},\cr
\tilde{{\rm{\textsl{m}}}}_{fi}^{ML} =
\frac{\kappa_i+\kappa_f}{L}
\int_0^{\infty}[g_i(x)f_f(x)+f_i(x)g_f(x)]\frac{dx}{x^{L-1}},
}
\label{eq:EME_NR}
\end{equation}
where $x=r/a_B$ [thus, in Eqs.~(\ref{eq:CS_NR})--(\ref{eq:EME_NR})
and below in the Dirac equations, the Bohr radius $a_B$ is used as
a natural parameter characterizing the size of the electron shell
of the Thorium atom (ion)]. Here the large, $g(x)$, and the small,
$f(x)$, components of the Dirac wave function are the solution of
the Dirac equations with the interaction potential $V(x)$:
%
% Eq. Dirac Equations
%
\begin{equation}
\left.
\begin{array}{ll}
g'(x)+\frac{1+\kappa}{x}g(x)-
\frac{1}{e^2}\left(\frac{{\cal{E}}}{m_e}+\frac{m_{\mu}}{m_e}-\frac{V(x)}{m_e}\right)f(x) =0,\\
f'(x)+\frac{1-\kappa}{x}f(x)+
\frac{1}{e^2}\left(\frac{{\cal{E}}}{m_e}-\frac{m_{\mu}}{m_e}-\frac{V(x)}{m_e}\right)g(x)
=0.
\end{array}
\right.
\label{eq:EqDirac}
\end{equation}
The function  $g(x)$ in Eq.~(\ref{eq:EqDirac}) is normalized at
$x\rightarrow\infty$ with the condition
$g(x)=\sin(pa_Bx+\varphi_{lj})/x$, where $\varphi_ {lj}$ is a
phase, $\kappa=l(l+1)-j(j+1)-1/4$.

The total potential of the nucleus and the electron shell was
built in the standard way, described in detail in
\cite{Tkalya-20-PRL,Tkalya-19-PRC-IC_Rydb}: the nucleus was taken
in the form of a uniformly charged ball of radius $R_0=1.2A^{1/3}$
fm, where $A$ is the atomic number, and the electron shell was
calculated within the DFT theory \cite{Nikolaev-15,Nikolaev-16}
through the self-consistent procedure. After that the potential
was obtained as a result of double integration of the electron
density.

\begin{figure}
 \includegraphics[angle=0,width=0.6\hsize,keepaspectratio]{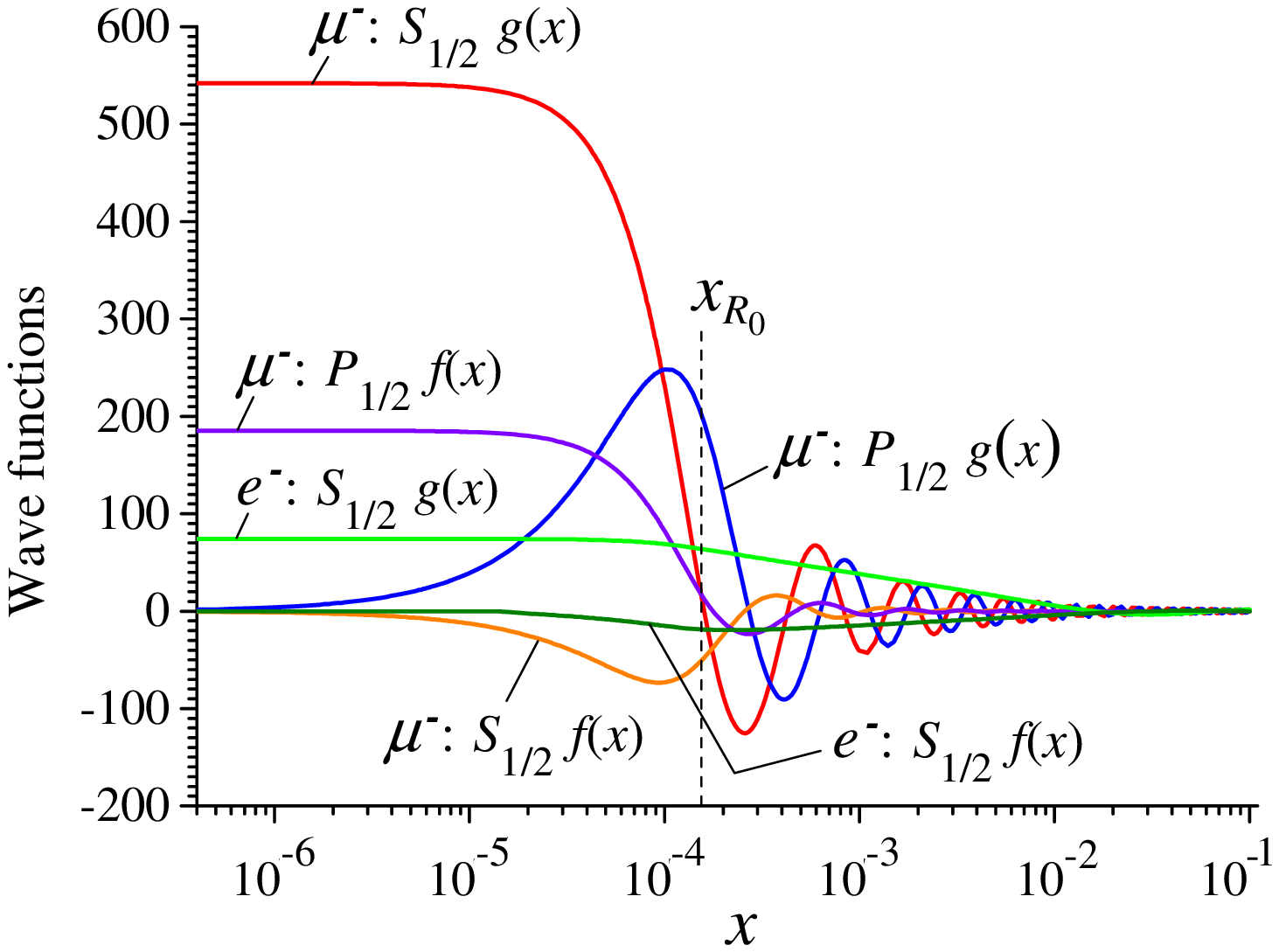}
 \caption{Wave functions of scattered muon and electron with the energy
 $E=10$~eV in the potential of  the Th atom.
 Here $x_{R_0}=R_0/a_B$ is a nuclear radius in the units of $a_B$.}
 \label{fig:WFs}
\end{figure}

The excitation by muons has its own characteristics. The wave
function of a scattering muon, in comparison with the wave
function of a scattered electron, has a much larger amplitude in
the region of the nucleus (see examples of wave functions in
Fig.~\ref{fig:WFs}). However, the wavelength of the muon is much
shorter than the wavelength of the electron and the muon wave
function oscillates more rapidly. Therefore, the muon matrix
element is formed in a much smaller region than the electron one.
Calculation shows that the muon matrix element reaches a plateau
already at $r\approx5R_0\approx 0.007a_B$, while the electron
matrix element is formed in the region $r\lesssim 0.1a_B$
\cite{Tkalya-20-PRL}. As a result of this compensation, the muon
and electron matrix elements in the case of the $M1$ excitation
turn out to be close in magnitude. As a consequence, the cross
sections for the magnetic dipole excitation of the $^{229m}$Th
isomer by the low-energy muons and electrons are comparable in
magnitude (see below in Fig.~\ref{fig:CS-Th-E2-Partial}).

%
%  Figure 2. Wave Functions of Scattering Muon and Electron. Th
%

The situation is different for the $E2$ excitation. The factor
$1/x^{L-1}$ in Eq.~(\ref{eq:EME_NR}) at $L=2$ enhances the
contribution to the matrix element from the region of small $x$.
But inside the nucleus, the amplitudes of the muon wave functions
by an order of magnitude exceed the amplitudes of the electron
wave functions. This explains a significant increase in the cross
section for the electroquadrupole excitation of the $^{229m}$Th
isomer by muons in comparison with electrons.

\section{Results of calculations}
\label{sec:Results}

\begin{figure}
 \includegraphics[angle=0,width=0.6\hsize,keepaspectratio]{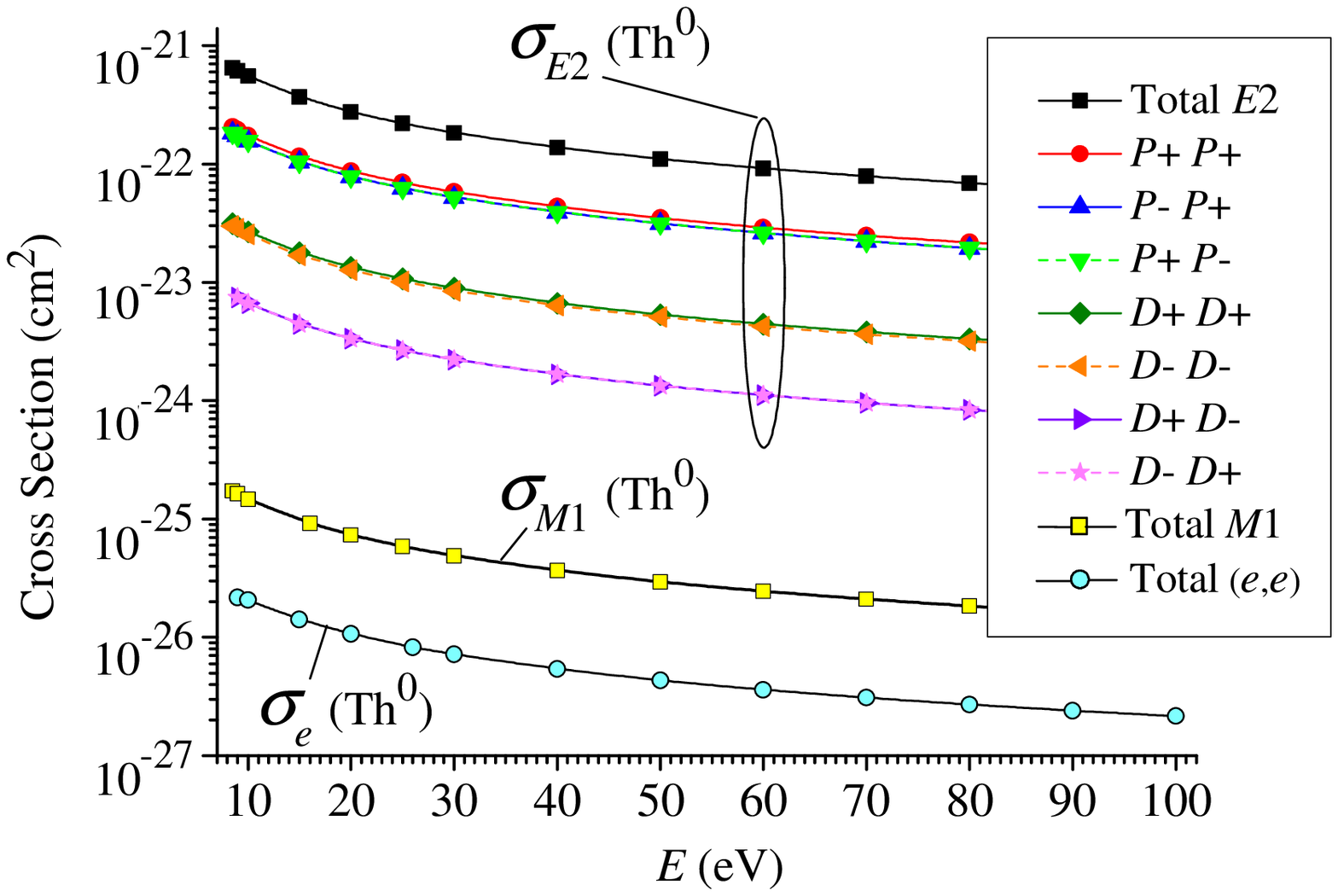}
 \caption{The total and partial $E2$ cross sections for the
 $^{229}$Th$(\mu^-,\mu^-)^{229m}$Th reaction for
 $B_{\rm{W.u.}}(E2;3/2^+\rightarrow 5/2^+)=11.7$. For comparison,
 the total $M1$ cross section for
 $B_{\rm{W.u.}}(M1;3/2^+\rightarrow 5/2^+)=0.031$ and the total
 $M1+E2$ electron cross section from \cite{Tkalya-20-PRL}
 $\sigma_e($Th$^0)$, are shown.}
 \label{fig:CS-Th-E2-Partial}
\end{figure}

\begin{figure}
 \includegraphics[angle=0,width=0.6\hsize,keepaspectratio]{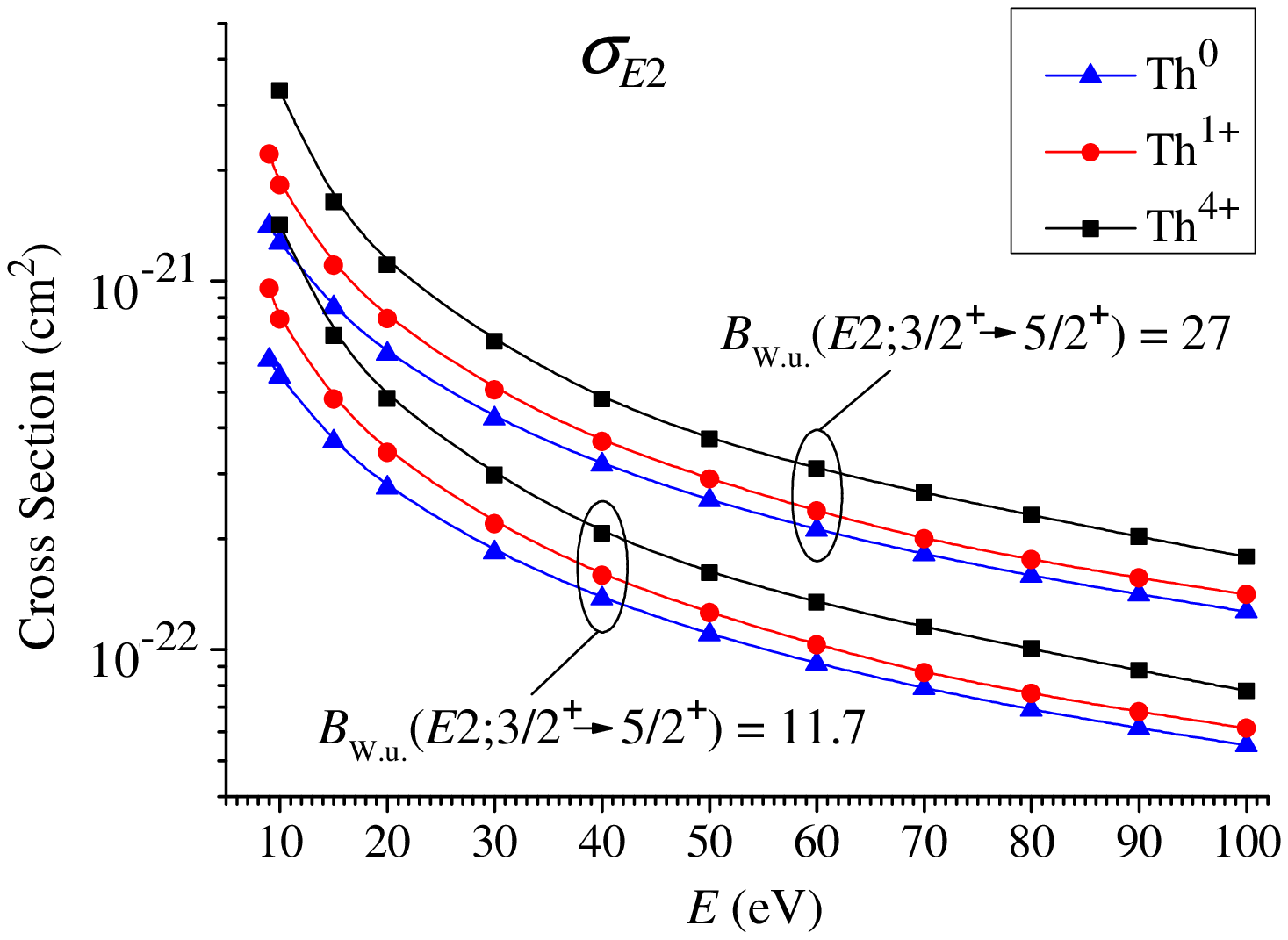}
 \caption{The cross sections of the $E2$ excitation of the $^{229m}$Th isomer
 by muons in the Thorium atom (Th$^0$) and Thorium ions Th$^+$ and Th$^{4+}$.}
 \label{fig:CS-E2-Total}
\end{figure}

The results of calculations of the $^{229m}$Th isomer excitation
cross section by muons are given in
Fig.~\ref{fig:CS-Th-E2-Partial} and Fig.~\ref{fig:CS-E2-Total}. As
expected, the cross section for the magnetic dipole excitation
turned out to lie close in magnitude (within an order of
magnitude) to the electron cross section from \cite{Tkalya-20-PRL}
and is approximately four orders of magnitude smaller than the
cross section for the electroquadrupole excitation,
Fig.~\ref{fig:CS-Th-E2-Partial}. This is an interesting and useful
observation, which allows us to extract the value of the reduced
probability of the $E2$ nuclear transition
$B_{\rm{W.u.}}(E2,5/2^+\rightarrow3/2^+)$ from the
$^{229}$Th$(\mu^-,\mu^-)^{229m}$Th reaction.

%
%  Figure 3. Partial E2 & Total M1 Cross Sections, Th
%
%

%
%  Figure 4. Total E2 Cross Sections: Th, Th+, Th4+
%

The total excitation cross sections of the $^{229m}$Th isomer for
two selected values of the reduced probability of the nuclear $E2$
transition $3/2^+\rightarrow 5/2^+$ are shown in
Fig.~\ref{fig:CS-E2-Total}. The first value,
$B_{\rm{W.u.}}(E2;3/2^+\rightarrow 5/2^+)=11.7$, is an average
over four experimental measurements
\cite{Bemis-88,Gulda-02,Barci-03,Ruchowska-06} of the $E2$
transitions between the $3/2^+[631]$ and $5/2^+[633]$ rotation
bands in the $^{229}$Th nucleus found with Alaga rules in
Ref.~\cite{Tkalya-15-PRC}). The second value,
$B_{\rm{W.u.}}(E2;3/2^+\rightarrow 5/2^+)=27$, was obtained
theoretically in Ref.~\cite{Minkov-17} as a result of the computer
calculations within the modern nuclear models.

Let us estimate the possible rate of excitation of the isomeric
nuclei by $\mu^-$. As it can be seen from the graphs in
Fig.~\ref{fig:CS-E2-Total}, for the problem under consideration,
muons with energies 10--100 eV are the most interesting, because
the excitation cross sections of the isomer have the values
$10^{-21}$--$10^{-22}$ cm$^2$ in this energy range.

A beam of $\mu^-$ slowed down to these energies was obtained in
\cite{Daniel-93}. Measurement of the stopping power showed that
the absorption of $\mu^-$ is very small in the range
$E=10$--100~eV \cite{Daniel-93}. This is an interesting result.
This implies that below the peak at around 10~keV, $\mu^-$ becomes
too slow to ionize atoms \cite{Nagamine-03} and stopping power is
induced by elastic collisions between the projectiles and the
target nucleus \cite{Garnir-80}. The isomer excitation reaction
considered in this work is an additional channel of the scattering
process on nuclei, and it will proceed in parallel with the
elastic collision.

An important question is how many negative muons will survive to
the required stage of deceleration. It is well known
\cite{Nagamine-03} that in the energy range $E\lesssim 5$~keV, the
dominant process is that in which the $\mu^-$ takes an atomic
orbit around a nucleus, yielding the muonic atom. A negative muon
captured on the atomic shell can excite the $^{229}$Th nucleus to
high-lying levels in a cascade of bound states transitions,
provided that there is a coincidence in energy and multipolarity
between the nuclear and muonic transitions \cite{Wheeler-49}.

In \cite{Daniel-93}, Kapton was used as a moderator for
decelerating negative muons to several eV. As a result, some of
the muons avoided being captured by its atoms, were slowed down to
energies of the eV range by passing through a Kapton layer of the
thickness $1.3\times10^{-4}$~cm. Then the beam of slow muons was
directed to the Au/Pd target, where the muons were gradually
stopped and captured by the target. In the range $E\leq 100$~eV
the stopping power was $\simeq1$~eV$\,$\AA$^{-1}$
\cite{Groom-01}. Therefore, the Au/Pd target thickness of about 7
nm was chosen so that the main part of the slow muons stopped and
were captured by the Au/Pd atoms. In accordance with the above
estimates, the optimal thorium target thickness $h$ should be in
the range 1--10~nm.

The fraction of slow muons $\xi$ was approximately
$10^{-4}$--$10^{-5}$ in respect to the particles of the initial
beam with $E=54$~MeV in \cite{Daniel-93}. Such a small value is
explained by the fact that the authors selected $\mu^-$ according
to the direction of motion and the value of energy. As a result, a
beam of muons moving in one direction and having a narrow energy
distribution, was produced. To excite $^{229}$Th nuclei, there is
no need to create a narrow directed monochromatic beam. Thus if we
take into account all muons passed through the Kapton (i.e., all
directions and all energies), then the value of $\xi$ will
increase by two--three orders of magnitude in comparison with the
quoted value of \cite{Daniel-93}.

The nuclear isomer $^{229m}$Th is registered using internal
conversion electrons with an efficiency $\eta$ close to 1
\cite{Wense-16,Seiferle-19}. For estimates, we also take the most
intense modern source of muons (muon factory), which gives
currents $j_{\mu}\approx 4\times 10^8$ s$^{-1}$ (i.e. 0.064~n$A$)
\cite{Cook-17}.

The rate of $\mu^-$ excitation of the isomeric nuclei in the
metallic target of the thickness $h=1$--10 nm from the
isotopically pure $^{229}$Th with the density of nuclei
$\rho_{\rm{Th}} =3\times10^{22}$ cm$^{-3}$ can be estimated by
the formula
$$
dN_{\rm{is}}/dt \approx{} \eta \rho_{\rm{Th}} h\, \xi\,
j_{\mu} \sigma_{E2} \simeq 10^0{\rm{-}}10^3\,\,
{\rm{s}}^{-1}.
$$
The first value $dN_{\rm{is}}/dt \approx 1$~s$^{-1}$ corresponds
to the most pessimistic scenario, while the second
($dN_{\rm{is}}/dt \approx 10^3$~s$^{-1}$) corresponds to the
most optimistic one.

Note that in the muonic scattering experiment, as well as in the
electron one, it is not necessary to tune precisely the muon
energy and to know the energy of the nuclear isomeric level
$E_{\rm{is}}$ with high accuracy, since the excitation process
is non-resonant.

\section{Conclusion}
\label{sec:Conclusion}

In conclusion, the cross section for the excitation of the
anomalous low-lying $3/2^+(8.10\pm0.17$~eV) level in the
$^{229}$Th nucleus by low-energy muons has been calculated. It is
shown that a) the muon cross section is four or five orders of
magnitude larger than the electron one, b) in the region $E\approx
10$~eV the muon cross section of the electroquadrupole excitation
reaches the value of $10^{- 21}$~cm$^2$ and is approximately four
orders of magnitude larger than that for the muon magnetodipole
excitation. The dominance of the Coulomb $E2$ excitation at the
muon scattering makes possible in future at the next generation of
the muon colliders to extract the value of the nuclear matrix
element of the $E2$ transition from the scattering data and test
some aspects of the existing theoretical models for the isomeric
state.

This research was supported by a grant of the Russian Science
Foundation (Project No 19-72-30014).

\end{document}